# Do Private Household Transfers to the Elderly Respond to Public Pension Benefits? Evidence from Rural China

Plamen Nikolov[abcd]☆     Alan Adelman[a]

**Abstract.** Ageing populations in developing countries have spurred the introduction of public pension programs to preserve the standard of living for the elderly. The often-overlooked mechanism of intergenerational transfers, however, can dampen these intended policy effects as adult children who make income contributions to their parents could adjust their behavior to changes in their parents' income. Exploiting a unique policy intervention in China, we examine using a difference-in-difference-in-differences (DDD) approach how a new pension program impacts inter vivos transfers. We show that pension benefits lower the propensity of receiving transfers from adult children in the context of a large middle-income country and we also estimate a small crowd-out effect. Taken together, these estimates fit the pattern of previous research in high-income countries, although our estimates of the crowd-out effect are significantly smaller than previous studies in both high-income and middle-income countries. (JEL D64, O15; O16; J14; J22; H55, R2)

We thank Matthew Bonci, Xu Wang and Jian Deng for outstanding research support with this project. We thank Subal Kumbhakar, Eric Edmonds, Susan Wolcott, Nusrat Jimi, Wei Xiao, Solomon Polachek, James MacKinnon and Morten Nielsen for constructive feedback and helpful comments.

☆Corresponding Author: Plamen Nikolov, Department of Economics, State University of New York (Binghamton), Department of Economics, 4400 Vestal Parkway East, Binghamton, NY 13902, USA. Email: pnikolov@binghamton.edu

[a] State University of New York (Binghamton)
[b] IZA Institute of Labor Economics
[c] Harvard Institute for Quantitative Social Science
[d] Global Labor Organization

# I. Introduction

How family members respond to public transfers can play an important role in assessing the efficiency and welfare impact of redistributive policies targeting the elderly. Of particular concern is the possibility that that public transfers induce a reduction in private family transfers, thereby dampening the redistributive effect of public transfers and resulting in an overall reduction of available savings, and thus welfare.[1] This scenario is exceptionally relevant to developing countries and rapidly growing economies, such as China, where such crowd-out effects can affect millions of elderly people, many of whom live in poverty. The World Bank (2001) estimates that expanding formal safety nets and public transfer programs will displace private transfers by 20 to 91 percent. Previous empirical studies that examine the magnitude of crowd-out effects on inter vivos private transfers provide mixed evidence or some support for positive small crowd-out effects (Cox 1987; Cox and Rank 1992; Lucas and Stark 1985; Cox et al. 1998; Cox and Jakubson 1995; McGarry and Schoeni 1995; Altonji et al. 1997), although these studies are largely based on the experience of high-income countries. Evidence on the existence and magnitude of crowd-out effects in emerging and middle-income economies, such as China, is extremely limited.[2]

In this paper, we examine the impact of China's New Rural Pension Scheme (NRPS) on inter vivos private transfers (from children to parents and vice versa), particularly among adults ages 60 and above.[3] The NRPS program was introduced in 2009 in response to rising demographic and old-age poverty concerns in the last decade (Liu and Sun 2016; Holzman, Robalino and Takayama 2009: 111–18).[4] In 2007, approximately 11 percent of China's population was age 60 or over, making up 21 percent of the world's elderly population (UN 2007). Similar to other developing countries, the Chinese government had faced additional pressing challenges: large rural and informal agriculture populations, high internal migration flows (Sabates-Wheeler and Koettl 2010), and weak institutions (Musalem and Ortiz 2011). The new program[5], a defined-contribution pension program, was made available to all rural residents

---

[1] Crowding out refers to the phenomenon when public sector involvement (or spending) reduces private forms of spending. Feldstein and Liebman (2002) review the early literature on various forms of crowding out.
[2] Cox et. al (2004) argues the magnitude of crowd-out effects in low and middle-income countries could be quite different than the ones detected in high-income countries because high-income countries have experienced a century of large public transfers.
[3] In each case, we limit our discussion to interhousehold transfers as the family transfer information collected by our survey source (China Health and Retirement Longitudinal Study) is only for parents and children not living together with the respondents.
[4] Feldstein and Liebman (2002) and Cutler and Johnson (2004) provide a historical overview of social pension programs in developed countries. Social pension programs are common in developed world and are primarily aimed to provide old-age insurance and aid consumption smoothing.
[5] Prior to the NRPS program, China had had experience with other public pension programs but these programs were largely decentralized. The country first introduced a rural pension scheme in 1986 by first piloting the rollout among rural residents. In the 1990s, the public pension system

over the age of 16. We examine the program's impacts on inter vivos transfers using various quasi-experimental techniques.[6,7] We use a large panel data set, from which we focus on a cohort of individuals for whom we have rich data on program participation and inter vivos transfers. Program eligibility was based on individual contributions and age. Our identification strategy relies on within-country variation in program implementation due to the staggered implementation of the policy across communities. We use this staggered implementation as a source of identifying variation to detect program impacts from pension participation between individuals living in newly integrated communities between 2011 and 2013, and individuals who were not offered similar program benefits. We use a difference-in-difference-in-differences (DDD) empirical strategy to estimate the intent-to-treat program impact of having access to pension benefits on individual-level outcomes. We also instrument the individual program participation with a community variable related to the program offer. Finally, because of potential right-censoring in our outcome variable, we use a censoring-data adjustment based on Cameron and Trivedi (2005).

Using data from 2009 through 2013 from the China Health and Retirement Longitudinal Study and the China Health and Nutrition Survey, we find statistically significant impact of the availability of NRPS benefits on lowering the probability of the incidence of receiving a transfer from adult children by 7.4 percentage points and lowered the actual amount of transfers by 10.4 percent[8].[9] We do not detect statistically significant evidence that the benefits' availability impacted transfers sent to children. The small amount of evidence provided by previous studies in developing countries support our findings that elderly pension benefits tend to crowd-out private transfers received.[10] Additionally, we find that receipt of benefits had a significant effect

---

was based on pension programs for state employees and the so-called Basic Old Age Insurance Scheme (BOISE). The program was scaled down in the late 1990s due to concerns about long-run sustainability in rural areas. In addition, the Ministry of Civil Affairs (MoCA) and the predecessor of the National Development and Reform Competition introduced various pilot pension programs in rural areas throughout the 1990s. These pilot programs aimed at establishing a financially sustainable pension scheme without long-term state subsidies. However, the Asian financial crisis and the lack of a general structure and oversight lead to gradual discontinuation of these pilot programs.

[6] Using data from two Chinese provinces (Guizhou and Shandong), Chen et al. (2018) study the impact of the NRPS program and specifically examine whether higher income for elderly parents leads to a higher rate of independent living, whether parents have better access to healthcare, whether adult children substitute away support care for their parents as a result of parental access to more healthcare, whether the NRPS program results in changes to monetary and non-monetary transfers between adult children and parents who are NRPS beneficiaries. The study relies on a fuzzy regression discontinuity design by exploiting the fact that the NRPS program kicks in benefits for individuals in rural areas when individuals reach the age 60 (the running variable in this RDD design). Although the study relies on a much more limited sample from two provinces, its results are suggestive of behavioral response for NRPS participants: the study reports that transfers between grandparents and grandchildren decrease in both Shandong and Guizhou

[7] Nikolov and Adelman (2019) examine the impact of the NRPS program on physical health, mental health, social interactions, and mediating individual behavioral inputs.

[8] Based on the average baseline transfer amount of 4,242 yuan (approximately 692 US dollars in 2013).

[9] Because of the right-censoring of the data for our main outcomes, our primary analysis relies on Tobit adjustment for the right-censoring.

[10] Cox et al. (2004), Juarez (2009), Gerardi and Tsai (2014) and Jung et al. (2016).

on the intensive and extensive margins of private transfers. NRPS benefit recipients were 10.4 percentage points (55.1 percent) less likely to receive transfers. Benefit receipt was associated with a decline in transfers of about 18.8 percent of the average transfer amount. Similarly, we did not find that benefit receipt affected private transfers sent to children. Finally, using Tobit analysis for the right-censored outcome measures, we investigate for the magnitude of the crowd-out effect. We find a crowd-out effect of -0.03 (imprecisely estimated[11]) suggesting that family members do offset their inter vivos transfers to elderly parents in response to pension benefits, but that substitution is extremely small in magnitude and much smaller than previously estimated using data from middle- and high-income countries.

Previous empirical studies examine the substitution between the availability of public saving options and subsequent intergenerational transfers and these studies find mixed results.[12] Studies that examine the relationship between public transfers and inter vivos transfers in the context of high-income countries (Cox 1987; Cox and Rank 1992; Lucas and Stark 1985; Cox et al. 1998; Cox and Jakubson 1995; McGarry and Schoeni 1995; Altonji et al. 1997) find evidence of substitution with estimated impacts on the probability of transfers that range from -0.01 to 0.015 (probability of decreased incidence of an inter vivos transfer).[13] In contrast to these findings, evidence from low-income and middle-income countries (Cox et al. 2004, Juarez 2009; Gerardi and Tsai 2014; Jung et al. 2016) also find evidence of substitution effects but the estimated substitution response is tends to much higher in the context of public programs in developing countries than in high-income countries. Therefore, the potential welfare consequences of the behavioral responses to newly introduced pension benefits in a large country are enormous. Furthermore, because of its fast-ageing population and a public saving option recently introduced to rural areas, China is a particularly compelling setting for this study.

This paper makes four important contributions to the existing empirical literature on inter vivos transfers and the possible crowd-out effects between various forms of saving. First, we contribute to the existing literature on intergenerational transfers in low and middle-income countries and we are among the first set of studies to shed light on how pension benefits influence inter vivos transfer in the context of a very large middle-income country.[14,15,16,17] In this study, we focus on China, which is the world's most populated country. The setting of this study makes the consequences of our results important from a welfare standpoint.[18,19] Ageing populations worldwide have prompted policy responses to alleviate old-age poverty. China's rural elderly have predominantly been without public assistance until the NRPS, making China a setting ripe to study the effects of an elderly pension program on private transfers. Thus, examining the potential behavioral response to the receipt of pension benefits in the context of the largest country has important implications for the design of public assistance programs in the context of other highly populated and developing countries. Our second contribution relates to our estimated program impact on the incidence of interhousehold transfers in response to pension benefits. All three studies that use data from middle-income countries do not explicitly test for program impacts on the incidence of new transfers and instead focus on the crowd-out effect.[20] We find a very small (relative to previous studies from middle-income countries) and statistically significant reduction in the incidence of inter vivos transfers to the elderly.[21] Third, previous studies that rely on data from middle-income countries (Juarez 2009, Jensen 2004 and Gibson, Olivia and Rozelle 2011) show a large crowd-out effect in response to pension benefits.

---

[11] We estimate a negative 0.03 effect size for the Marginal effect (i.e., E[T | T>0]) but the effect size estimate is imprecisely estimated because of large standard errors.

[12] Existing economic theory on the relationship between private and public transfers focuses on two important features guiding family member exchanges -- altruism and self-interested exchange (Cox 1987). These two features lead to important theoretical predictions. Assuming altruistic family transfers, the introduction of public transfer programs will offset private transfers (Becker 1974; Barro 1974). If exchange motives guide private transfers, public transfers will not necessarily undermine private transfers (Bernheim et al. 1986; Cox 1987; Cox and Jimenez 1990; Morduch 1999; Cox and Fafchamps 2007). However, which of these theoretical predictions play out in practice across various settings remains an empirical issue.

[13] In the U.S., for example, the estimated decrease in probability of private transfers is 0.002 (based on Altonji et al. 1997). The estimated impacts on the actual dollar amount decrease of private transfers received per dollar received in income ranges from 3 cents (McGarry and Schoeni 1995) to 13 cents (Altonji et al. 1997), suggesting that crowding out from government programs in the U.S. is negligible.

---

[14] Our study is closely related to a strand of literature (in high-income countries) that examines the substitution between private wealth and public pension provisions in high income countries. Using data from the U.S., Diamond and Hausman (1984), Hubbard (1986) and King and Dicks-Mireaux (1982) examine the degree of substitution between private wealth and public pensions. They find small offsets between the two. Similarly, Brugiavini (1987) and Jappelli (1995) obtain large estimates, using data from Italy, of the degree of substitutability between private wealth and public pension provisions.

[15] Related empirical literature, also in high-income countries, examines the substitution between various savings devices. Poterba, Venti and Wise (1996) and Engen, Gale and Scholz (1996) review the public finance literature in high-income countries. Some recent studies in this literature (e.g., Gelber 2011) present evidence that increases in IRA or 401(k) savings represent increases in total saving, while others (e.g., Benjamin (2003), Engelhardt and Kumar (2007), Chetty et al. 2014) find that much of the increase in 401(k) savings represents substitution from other accounts. Although some of the difference between the results of these studies likely stems from differences in econometric assumptions, the variation that drives changes in contributions to 401(k)'s could also explain the differences in results. For instance, increases in 401(k) contributions by employers may generate less crowd-out than tax incentives or programs that require active individual choice, an idea foreshadowed in early work by Cagan (1965) and Green (1981).

[16] Juarez (2009), Jensen (2004) and Gibson, Olivia and Rozelle (2011) study the effect of public programs on private transfers in the context of other middle-income countries (respectively in Mexico, South Africa and Vietnam).

[17] Gerardi and Tsai (2014) and Jung et al. (2015) examine the effect of pension benefits on interhousehold allocation decisions but in the context of a high-income country.

[18] China's age dependency ratio has been rising rapidly in the 2000's, more so in rural areas, and by 2030 should resemble Japan's ageing rate as of the last 30 years (Cai et al. 2012).

[19] From 2013-2016, China has experienced one of the largest changes in age dependency ratio (2.16 PP change). Much larger than developing countries in the Middle East (0.56), East Asia and Pacific (1.24) and Latin America (-1.29). From 2013-2016, China has been ageing faster than the European Union (2.11) and the U.S. (1.33). This is based on the author's calculations using the age dependency ratio (%) provided by https://data.worldbank.org/indicator/SP.POP.DPND. The age dependency ratio is defined as the ratio of dependents, people younger than 15 or older than 64, to the working-age population, those ages 15-64.

[20] Using data from high-income countries, Gerardi and Tsai (2014), Jung et al. (2015) and Fan (2010) do examine for program impacts on the incidence of interhousehold transfers and find large effects (i.e., reduction of probability by 0.4) relative to our estimates.

[21] Cox and Jakubson (1995) and McGarry and Schoeni (1995) find a similar pattern but much larger estimates in high-income countries.

Although our estimates are in line with the overall substitution pattern of the estimated crowd-out effect from middle-income countries, we find much smaller crowd-out effects (imprecisely estimated) relative to previous empirical studies using data in middle-income countries. Our fourth contribution relates to our large sample size relative to studies from low- and middle-income countries. We rely on data from almost 12000 households. Finally, we examine for heterogeneous effects, by household income status, of the introduction of the public program on the incidence and the crowd-out effect of private transfers. Although we don't detect differential impacts when we formally test for equality of the crowd-out effect by household income status, we do detect a larger decrease in the incidence of the private transfers among the high-income households.

This paper is organized as follows. Section 2 provides background of rural pension programs in China and on the NRPS. Section 3 presents a conceptual framework for intergenerational transfers. Section 4 summarize the data and estimating sample. Section 5 presents the identification strategy. Section 6 discusses the results. Section 7 presents various robustness checks. Section 8 concludes the paper.

## II. China's New Rural Pension Scheme

### A. Introduction of the NRPS Program

In the last three decades, the Chinese government gradually took on the responsibility to alleviate poverty, particularly the risk of old-age poverty. To this end, the country introduced a rural pension scheme in 1986 by first piloting the rollout among rural residents before expanding further coverage. The financing of the program relied on voluntary individual contributions with matching from the local government. From the nineties onwards, the public pension system was based on pension programs for then state enterprise employees and the then called Basic Old Age Insurance Scheme (BOISE), to the introduction of a New Rural Pension Scheme in 2009 (Liu and Sun 2016). For the first time, compulsory coverage quotas for both urban and rural systems were included. By the end of 1998, two-thirds of the rural communities were covered, or 2123 communities in 31 provinces. However, a combination of poor governance, unsound local operations and inflationary pressures brought about by the Asian financial crisis in 1997, halted the rural pension expansion. The program was scaled down in 1999, following concerns about its long-run sustainability in rural areas. Pension coverage fell from 80.25 million participants in 1998 to mid-50 million among rural participants in 2007.[22]

The New Rural Pension Scheme (NRPS) launched in 2009. The program aimed to achieve full geographic coverage by 2020 (Dorfman et al. 2013; Cai, Giles, O'Keefe and Wang 2012). Program features encompass a basic flat pension financed by the central government, individual contributions and minimum matching by local governments. In the beginning, the program covered 23 percent of communities at the end of 2010, and over 60 percent of communities by early 2012. Figure 1 shows program coverage expansion over time.

[Figure 1 about here]

Total participation grew to 326 million from 2009 to the end of 2011 (Quan 2012) and over fifty percent of rural residents participated in the NRPS by the end of 2011. The program began with 320 pilot counties in 2009. The program gradually expanded and covered 838 counties by 2010 and by 2012 it reached almost all 2853 counties. Four important factors likely account for this program expansion. First, China's initial localized dedication to rural pension reform. Second, its high economic growth rate from 2009-2011 likely played an important role. Third, an increase in demand for the basic monthly benefit in untargeted areas generated by the initially unequal geographic coverage. Last, the rural pension expansion was a key element in domestic policy discourse in 2012, this political debate further intensifying interest in the pension program.

**B. Program Eligibility and Benefits**

The NRPS is available to all rural residents over the age of 16 not enrolled in an urban pension program. The program was introduced in rural administrative districts called *Hukou*.[23]

---

[22] The pension system faced continuing challenges in the early-2000s, especially in rural areas. The rural residents had widespread mistrust in this pension setup and the system failed to establish itself in the rural areas. Although program participation reached 80 million in 1999 (approximately 11 percent of the total rural population at that time), participation steadily declined thereafter as more and more residents withdrew their payments from their individual accounts. Furthermore, program participation favored wealthier regions and poor provinces failed to make their matching contributions. The program witnessed a resurgence from 2003 as interest grew and individual participation rapidly increased. More than 300 communities and 25 provinces introduced program benefits by the end of 2008 (Dorfman et al. 2013). Reformist elites under the Hu-Wen in the late 2000s sought to create a novel basic and non-contributory pension for individuals not covered by any social insurance program. A 2008 pilot project in the city of Baoji (Shaanxi province) stirred attention for a similar program implemented on a nationwide scale. In the Baoji pilot, local administration introduced a pension scheme covering rural residents. The pension scheme was funded through local tax revenues. In addition to the basic pension, residents were encouraged to participate in a separately funded pension plan, subsidized by the local administration. Hence, the Baoji rural individuals were covered by a two-part pension plan. This "Baoji model" (Qing, 2009) served as a template for the new national program about to be introduced nationwide.

[23] The *hukou* system is a governmental household registration system in China. The plan was implemented by the Chinese Communist Party as a classification system to keep record of all Chinese as either a rural or urban citizen. The Chinese government tied all social benefits (e.g., healthcare, education, social security, working rights) to a person's local government. The system is called "huji" but it's commonly known by the name of the records "Hukou." The Hukou System was implemented in 1958 and it still in place today. In China, the hukou system is also

Program participation is voluntary, and those who contribute for at least 15 years are eligible to receive benefits upon reaching the age of 60. Rural residents, who are over the age of 60 at the start of the program, are eligible to receive the basic monthly benefit of 55 RMB if their children already contribute to the pension scheme.[24] Individual benefits are calculated following the "*139 Rule*". The administrator takes the accumulated balance in an individual's account and divides it by 139. [25]

Individual contributions are voluntary and range annually from 100 to 500 Chinese RMB (approximately 15 to 77 US dollars). Based on 2009 survey data, the mean participant contribution was 100 RMB (Dorfman et al. 2013). The local governments are required to match 30 RMB annually per contribution. Participants between the ages of 45-60, with less than 15 years of contributions, are encouraged to increase monthly payments to cover the absence of contribution over the work cycle. Rural residents who opt to participate in the program need to register at their local village government office, the lowest level of the government hierarchy. Payments are made at that office as a lumpsum cash payment throughout the year. At the end of each contribution period (i.e., a month), each village-level government aggregates payments to it and transfers payment the next level (a township-level government), which then transfers the contributions the next level (i.e., county). Although there have been some policy discussions for the funds to be managed at the province level, currently the rural pension funds are generally managed by the county-level government. In general, the contributions are deposited in banks and administrative expenses are picked up by the local governments.

## III. Conceptual Framework

We provide a general framework to motivate how household members decide on inter vivos transfers.[26] Consider a simple model that assumes: (1) a parent and a child, (2) a utility

---

effectively limiting where a person can live, especially if one is born into a rural hukou – attempting to change to a more attractive residence or to an urban hukou can be extremely difficult, if not impossible.

[24] The central government fully subsidizes the basic pension in Central and Western provinces and splits the cost with local governments in Eastern provinces (Cai, Giles, O'Keefe and Wang 2012).

[25] For example, the amount of the NRPS pension was constructed as follows: the base of the amount was 55 plus the actual individual contribution divided by 139. The "139" is the average expected life expectancy (in months) at age sixty. On average, the pension benefits of the NRPS pension program are approximately 70 RMB per month. Finally, the pension plan exhibits large regional disparities in terms of benefits. Although the central government has set a minimum of 55 RMB (approximately 9 US dollars) per month as a minimum benefit, the local administrations can supplement individual benefits depending on their fiscal capacity or on the local cost of living. Therefore, the individual benefit levels can vary tremendously across regions.

[26] Cai et al. (2006) theoretically model the extent to which private transfers respond to the failure of China's city-based pension schemes. The study models whether altruistically motivated private transfers insure retirees against low income in old age. Furthermore, it empirically focuses on the extent to which intergenerational altruism can make up for formal sources of support.

normal good, (3) a parent with perfect information, (4) an individual good of monetary transfers, (5) a single period, (6) altruistic family members, and (7) exogenous income sources. The transfers are between two types of individuals: a donor (child) and a recipient (parent). The donor (e.g., an adult child) makes transfers $F$ to a family recipient (e.g., the child's parent). The parent's utility is denoted as $U_p$ and the adult child's utility is denoted $U_c$. Private consumption by the parent and adult child are denoted by $C_p$ and $C_c$, respectively. The relationship between the utility functions of the adult child and the parent can be expressed as $U_c = U(C_c, V(C_p))$. The child chooses $C_c$ to maximize his utility:

$$\max_{C_c} U(C_c, V(C_p)) \quad (1)$$

subject to

$$F \geq 0 \quad (2)$$
$$C_p = I_p + \tau_p + F \leq Y_p \quad (3)$$
$$C_c = I_c - \tau_c - F \leq Y_c \quad (4)$$

$U_c > 0$. $U_v$ captures the intensity of the altruism (e.g., caring parameter) with $0 < U_v < 1$. We assume three income sources for the parent: financial support from adult children ($F$), net public welfare support ($\tau_p$) and other types of incomes ($I_p$). $C_p$ is the consumption of the parent and is given by constraint (3). The consumption function of the child includes the feature that requires him to subsidize net public transfers ($-\tau_p$). The adult child also has other sources of income ($Y_c$). The adult child's consumption function is $C_c$, as specified by (4).

Because the adult child's income taxes subsidize the transfer, the public welfare support ($\tau_p$) received by older people is de facto a net public transfer ($\tau_c$). The implication of this relationship is that the total income of the family remains unchanged as $\tau_p = -\tau_c$. The child's objective function including the transfers now is: $U_c = (I_c - \tau_c - F, V(I_c + \tau_p + F))$ with the child choosing $F \geq 0$ subject to the three constraints above.[27]

The FOC yields $-U_c + U_v U_p = 0$. Cox (1987) shows that the interior solution to the household maximization problem where household members choose F is: $\frac{\partial F}{\partial Y_c} - \frac{\partial F}{\partial Y_p} = 1$ (also known as the income pooling result).

---

[27] Laferrère and Wolff (2006) overview carefully the rich tradition of various family transfer models based on what they call various "pillars" of the pure altruism model.

Two scenarios emerge out this framework. In the first scenario, the altruistic child can subsequently reduce a dollar of familial transfers previously given to his parent when he provides one dollar to support the public welfare system.[28] In this setup, a complete crowding-out occurs – in other words, a dollar increase of public transfers to the parent will lead to a dollar decrease of private transfers from the child to the parent. In the second scenario, the adult child may reduce his support by less than the amount of public transfers received by his parent, if the adult child bears no specific tax burden for the new welfare program (*i.e.*, $\tau_p \neq -\tau_c$). In this case, the public transfers become windfall benefits to the family. A partial crowding-out occurs instead of a complete crowding-out.

## IV. Data

### A. Survey Data

*China Health and Retirement Longitudinal Study.* Our primary data source is The China Health and Retirement Longitudinal Study (CHARLS), from which we draw data on retirement status, pension program access and benefits, family transfers, as well as individual-level and community-level socio-economic information. The CHARLS is a nationally representative survey that samples individuals 45 years of age or older and their spouses. The CHARLS collects data on demographic information, family structure, subjective and objective health status, health care use, pensions and retirement, work, household wealth, income, and consumption. The sample consists of 17708 individuals living in 10287 households in 450 villages/urban communities in 150 counties across 28 of China's 30 provinces, excluding Tibet. Figure 2 shows the survey coverage map. Basic information on education, gender, age, household size, and marital status was collected at the individual and household-level. The survey response rate was over 80 percent (94 percent in rural areas and 69 percent in urban areas).[29,30]

---

[28] This type of outcome is known as the Ricardian Equivalence. According to this theory, when a fiscal policy is implemented, altruistic family members will redistribute resources among themselves to neutralize the effect of the policy.
[29] The sampling process occurred in three stages. First, all community-level units were stratified into 8 regions, by rural and urban communities and by community/district GDP per capita. After this step, 150 communities were randomly chosen using probabilities proportional to size (PPS). Within the 150 communities, three primary sampling units (PSUs) were randomly selected using the same PPS method. Households were selected for an interview if a member of the household was greater than or equal to 39 years of age. If the spouse of the main interview respondent was present, then the spouse was also selected for an interview. As a result of this process, 17708 individuals from 10069 households in 450 communities were selected for survey interviews. The individual and household surveys were administered between June 2011 and March 2012 at the respondents' home using CAPI technology. The surveys collected information on the household income, expenditures, and assets.
[30] Household size in the CHARLS survey includes the number of household residents including the respondent and/or his or her spouse.

Once the 450 PSUs were selected, age-eligible households were interviewed. The 2011 baseline wave interviewed 10,257 households with 18,245 respondents aged 45 and over.[31] The follow-up 2013 wave covered 10,979 households (or 19,666 respondents). The follow-up wave had a high response rate of 88.6 percent of the original participants and 89.6 percent of original households.[32] The 2013 CHARLS wave added 2,053 new households with 3,507 individuals. The 2016 harmonized CHARLS dataset merged all modules in the 2011 and 2013 waves.

We present an overview of baseline characteristics (based on the CHARLS sample) in Table 1. Among the eligible sample of participants and non-participants, respectively 70 percent and 69 percent were employed in the baseline. About three-fourths of the sample worked in agriculture: 72 percent among participants and non-participants alike. The rural sample reported low levels of educational attainment – approximately 48 percent of participants and 46 percent of non-participants reported having completed at least the secondary level of education. The average non-transfer household income is 25,665 yuan, of which additional household member income makes up 35 percent. Although a balanced sample is not a necessary identifying assumption for our empirical strategy, our sample is balanced across most of the characteristics, including income, earnings, consumption, and hours worked. As a result, the differences between the two groups is likely be stable over time and any changes in treatment exposure is less likely to be associated with changes in the distribution of covariates. We formally test if the difference in means is statistically different from zero for each of the listed variables. It is reassuring that most of the variable means between treated and non-treated areas are statistically similar.

[Table 1 about here]

In the baseline, participants were more likely to receive transfers than non-participants. Around 43 percent of participants were receiving transfers from children, compared to the 36 percent of non-participants. Both groups had around 16-18 percent of respondents that transferred money to their children. In monetary terms, participants gave, on average, 813 yuan and received 2051 yuan. Whereas, non-participants gave 765 yuan and received 1534 yuan.

---

[31] Initially, 19,081 households were sampled and 12,740 had age-eligible members, of which, 10,257 responded.
[32] The number of tracked respondents in the follow-up wave was 16,159 (from the original 18,245 respondents); 9,185 of the original 10,257 households participated in the follow-up survey.

*China Health and Nutrition Survey*. We additionally supplement the CHARLS data with survey data from The China Health and Nutrition Survey (CHNS). We use auxiliary CHNS data on study outcomes because CHNS spans years prior to program implementation and it enables us to draw on outcome information prior to 2009 (i.e., the year of the program introduction) on family transfers, individual-level, and district-level socio-economic information. Using CHNS survey data prior to the introduction of the NRPS enables us to test important identifying assumptions. The CHNS is a longitudinal survey that covered about 19,000 individuals in 15 provinces spanning 216 PSUs.[33] Figure 2 reports the coverage map for the CHNS survey. The CHNS dataset is one of the few panel datasets that has collected data on individuals every two or four years for several decades. It has collected a wide range of variables at the individual, household, and community level, including: income, employment, health and nutrition, consumption, water sources, sanitation, demographics, and access to improved roads.

[Figure 2 about here]

The survey started in 1989 and covers the years 1989, 1991, 1993, 1997, 2000, 2004, 2006, 2009, and 2011 with the intent to provide data on how certain economic and social factors affect individual health and nutrition. Subsequently, the CHNS modules include food choice, nutritional intake, health behaviors, physical activities, work activities, time usage, and nutritional status. The sample selection process is similar to the multi-stage random selection process used by the CHARLS. First, communities were stratified by income level, followed by a weighting scheme that selected four communities from each province (CHNS Research Team 2010).[34] Individual respondents were asked about family transfers, that is, the amount of yuan received in the last year from children, parents, friends, and relatives.

## B. Program Participation and Survey Measures of Inter-Household Transfers

To measure NRPS program participation, we rely on the CHARLS survey and how it captures one's NRPS program participation. In the CHARLS, respondents are specifically asked

---

[33] The survey covered the following provinces (also presented in Figure 2): Bejing, Chongqing, Guangxi, Heilongjiang, Henan, Hubei, Hunan, Jiangsu, Liaoning, Shaanxi, Shandong, Shanghai, Yunnan, and Zhejiang.

[34] From the 2004 survey onwards, all questions related to individual activities, lifestyles, health status, demographic status, body shape, and mass media exposure, etc., were added to the individual questionnaires. The individual questionnaires are split into two parts: one is for adults aged 18 and older and the second is for children under age 18. Children aged 6 and above and all adults provide their time allocation on household and physical activities, as well as food and beverage consumption. Additional information is collected on smoking status, alcohol consumption, diet, and physical activity for adults and children 12 years and older. Adolescents aged 12 and older and women under age 52 with children aged 6 to 18 living in the household were asked additional questions related to mass-media exposure. For adults aged 55 and older, daily living activities and memory test scores were provided

"Do you participate in the New Rural Social Pension Insurance program?" Using this question, we can identify which individuals participated in this pension program. Additionally, the questionnaire elicits information on the timing of NRPS benefits, as well as the monthly benefit amount (reported in yuan).[35] NRPS participants are classified as benefit recipients once they begin receiving income benefits at the age of 60, or older. Individuals aged 60 and over at the time of NRPS implementation can receive benefits if their living children participate in a pension scheme. Since the NRPS is a voluntary program, we classify non-participants as eligible individuals who choose not to participate in the pension program.

We construct an eligible sample based on the general program guidelines. Mainly, we drop observations with an urban Hukou status because these individuals are ineligible to participate in NRPS. For individuals living in eligible districts, we drop individuals in the baseline who are over the age of 60 and without living children. We also drop urban pension participants with a rural *Hukou*.[36] In the eligible sample, we can directly observe NRPS participants and non-participants. Our eligible sample consists of 15990 individuals from 429 communities in 121 cities across 28 provinces.

The CHARLS has a comprehensive coverage of family transfers. The survey asks respondents to report transfers to/from children, parents, and other relatives. Since financial transfers can represent multiple aspects of family support, the questionnaire distinguishes between the type of support. Specifically, respondents report the amount of monetary and in-kind support given or received. These financial transfers between family members occur on a regular and non-regular basis. For regular transfers, the respondents report a monthly, quarterly, or half-year interval. This information is used to construct the annual amount transfers given/received between the respondent and adult children, parents and other relatives.[37]

## V. Empirical Strategy

---

[35] Since benefits are reported in monthly terms, we adjust the variable to an annual benefit amount.
[36] Both groups of individuals are not eligible to receive NRPS monthly benefits.
[37] There are four types of family transfers that respondents report in the CHARLS. Monetary support can occur on a regular basis and may be used to cover living expenses, water, electricity, telephone expenses, loans, or rents. Non-regular monetary support can cover infrequent events such as festivals, birthdays, weddings, and funerals. On the other hand, in-kind transfers are material transfers in the form of food, clothing, etc. These can occur on a regular and irregular basis. The annual transfer amounts are constructed using a two-step process. First, the regular monetary and in-kind transfers are imputed based on the reported intervals. Second, we sum the imputed regular transfers and the non-regular transfers to get the transfer amount in the past year.

*Difference-in-Difference-in-Differences Approach*. Our identification strategy relies on within-country variation in the pension policy implementation due to the staggered rollout of the policy across communities.[38] We use this staggered implementation over the years 2011 and 2013 as a source of identifying variation to detect the program impacts from pension participation between individuals living in newly integrated communities and individuals who were not offered similar program benefits.

Our first estimation approach is a difference-in-difference-in-differences (DDD) strategy to estimate the program impact of having access to pension benefits on individual-level outcomes. We begin by constructing $OfferNRPS_{ct}$ for communities that offer the NRPS at time $t$.[39] To estimate the program's intent-to-treat effect, we use a two-way fixed effect DDD model based on Hansen (2007) and Bertrand, Duflo, and Mullainathan (2004). However, because program benefits are only available to individuals aged 60 and above, we interact the program offer with an age indicator, a binary indicator for whether an individual is aged 60 or over, following (Katz 1996; Gruber 1994; Rossin 2011). This approach is more efficient and leads to more precise estimates of the program effect.

(1) $$T_{ict} = \beta_0 + \beta_1(OfferNRPS_{ct} \times Above60_{ict}) + \beta_2 Above60_{ict} + \beta_3 X_{ict} + \phi_c + \mu_t + \phi_c \times \mu_t + \varepsilon_{ict}$$

$T_{ict}$ is the outcome in our analyses: the extensive and intensive margins of transfers to and from adult children. $Above60_{ict}$ is equal to 1 if the respondent is aged 60 and over. $\beta_1$ in (1) is the coefficient of interest (DDD estimator), which captures the estimate of the average effect of the program on eligible individuals over the age of 60. The coefficient captures the average effect of program availability on the outcomes of those above 60 years who live in a treated community, regardless of whether they decided to participate in the program -- *intent-to-treat* (ITT) effect. $X_{ict}$, is a vector of individual-level controls.[40] $\phi_c$ and $\mu_t$ are community-level and time fixed effects. Community-level fixed effects allow us to account for time-invariant characteristics that affect the likelihood of program availability in the community. Time fixed effects account for community-specific characteristics that could directly influence the health-related outcomes. Additionally, we use community-time fixed effects, $\phi_c \times \mu_t$, to control for community differences during the implementation of NRPS.[41]

---

[38] 23 percent of program communities were covered by the end of 2010 and over 60 percent by 2012 (Dorfman et al. 2013). Figure 1 shows newly treated communities between 2011 and 2013.
[39] Given data limitations, we construct $OfferNRPS_{ct}$ based on individual-level data. If no individuals indicate having NRPS at time $t$ in community $c$, then $OfferNRPS_{ct}$ equals 0. If at least one-person reports participating in NRPS, then $OfferNRPS_{ct}$ is set to 1.
[40] In each specification, we account for education, gender, household size and marital status. Inter vivos transfers tend to be strongly related to the household size and various socio-economic characteristics of the recipient (McGarry and Schoeni 1995; McGarry 1999).

For (1) to yield unbiased estimates of program impact, annual variation in the NRPS program offer across communities should be unrelated to any other observed or unobserved community-specific shocks. The DDD design we use relies on an identifying assumption (common trends assumption) that the important unmeasured variables are either time-invariant group attributes or time-varying factors that are group invariant. Together, these restrictions imply that the time series of outcomes in each group should differ by a fixed amount in every period and should exhibit a common set of period-specific changes. In other words, the identification assumption implies that treatment communities that provide NRPS program benefits would otherwise have changed in a manner similar, on average, to the control communities that did not happen to provide NRPS program benefits. Even though the identification assumption cannot be tested directly, we can examine to some extent whether the two groups (treated and non-treated) exhibit parallel trends in the outcomes prior to 2009 when the program was introduced.

Even though the identification assumption cannot be directly tested, we can somewhat examine whether the two groups, treated areas that offered program benefits and non-treated areas that did not offer program benefits, exhibit parallel trends (i.e., the average change in the non-treated areas represents the counterfactual change in the treated areas if there were no treatment) in the study outcomes prior to 2011 (i.e., the first year for which we have data). The CHARLS is a relatively new dataset that began in 2011 and serves as the baseline year in this study.

Although CHARLS does not have data prior to the NRPS intervention, we can use data from an alternative data source, the CHNS, to analyze data on pre-trends related to the identification assumption regarding parallel trends. The CHNS dataset is particularly suitable as it covers the CHARLS sampling areas and it elicits information from respondents regarding inter vivos transfers received from children. The main challenge in this particular analysis of the identifying assumption (for potential pre-trends between "treated" and "non-treated" units) is that the community identifiers or the geographic level variables do not match between the two surveys. Therefore, for this exercise, we had to re-define "treated" and "control" units at the province level (as opposed to the community level) in order to make use of geographic variables available in the CHNS survey and then proceed with testing for the parallel trends assumption with CHNS data for the period 2004 to 2009 but on the province level. It is important to underscore that we define a treated province in the CHNS data for the period 2004 to 2009 based on baseline data from the CHARLS for the percent of NRPS participating communities within a given province. Specifically, we define a "treated" province by using the percent of communities/"community IDs" that indicated (in the CHARLS survey) that they participated in the NRPS program within a given province to capture "treatment intensity" of a province and we then define the province's "treatment status" with a binary definition, i.e., "treated" or "not treated" province. Using this definition of a "treated" province can be used in the CHNS survey. However, it is important to acknowledge that we rely on a threshold choice regarding the percent of NRPS participating communities within a province to define a province with a "treatment" status.[42,43] We implement a formal test of the common trends assumption following Autor (2003). Using the CHNS survey data, prior to 2011, on several private transfer variables from the 2004, 2006 and 2009 waves, we can estimate the following specification:

(2) $$T_{ict} = \beta_0 + \beta_{-3} D_{ct} + \beta_{-1} D_{ct} + \phi_c + \mu_t + \varepsilon_{ict}$$

---

[41] The community-time fixed effects can account for community-level unobservable characteristics that could affect the study outcomes. Therefore, our DDD estimate is net of community specific time-varying characteristics.

$T_{ict}$ is the transfer outcome of interest, and $\phi_c$ and $\mu_t$ are community-level and time fixed effects. The coefficients on $D_{ct}$ estimate the interaction of time-period dummy variables and the treatment indicator for the first pre-treatment period and last pre-treatment period.[44] Appendix Table A.1 reports the results. From the results in Appendix Table A.1, we fail reject the null hypothesis that trends in the outcomes between treatment and control areas are the same, as $\beta_{-3}$ and $\beta_{-1}$ are insignificant.[45]

---

[42] We define a province as "treated" if it had more than 67 percent coverage rate based on the percent of NRPS participating communities within it at the baseline. We conducted additional sensitivity analyses based on alternative choices regarding the coverage rate. In these additional sensitivity analyses, we varied the threshold choice so that we could impose lower and higher thresholds (e.g., 40 percent coverage, 50 percent coverage and 70 percent coverage) to define a given province as a "treatment" or "control" province; the pattern of these sensitivity analyses remained robust to coverage rate choice and we do not detect any evidence of violation of the parallel trends assumption based on alternative choices for coverage rate within a province.
[43] The CHNS doesn't sample from the same communities/villages as the CHARLS, so we rely on our definition of treated and control provinces based on the CHARLS to test data in the CHNS.
[44] We omit the interaction for the second pre-treatment period.

*Two-Stage Least Squares Approach.* We further address the possibility that time-varying unobservable characteristics at the community level could bias estimates of $\beta_1$ in (1). To address potential endogenous selection of program targeting (i.e., whether a community offers the NRPS program or not), we augment the DDD analysis by additionally instrumenting the program offer at the community level. We thus re-estimate (1) using a combination of the DDD estimation along with an instrumental variable strategy. The provision of pension benefits in a community may have been a function of its dependency ratio (or factors related to it) in 2009. Therefore, the estimate could potentially confound the effect of the program with mean reversion that would have taken place even in its absence. Thus, we use specifications that control for the interactions between cohort dummies and time periods. We use $OfferNRPS_{ct}$ to instrument for individual participation in the NRPS to augment the DDD estimation, similar to the approach employed by Nunn & Qian (2014). $OfferNRPS_{ct}$ equals 0 if no individuals participate in the NRPS and equals 1 if the community has at least 1 participant.[46] We estimate:

(3) $\quad T_{ict} = \beta_0 + \beta_1(\widehat{NRPSReceipt}_{ict} \times Above60_{ict}) + \beta_2 Above60_{ict}$
$\quad\quad\quad\quad + \beta_3 X_{ict} + \phi_c + \mu_t + \phi_c \times \mu_t + \varepsilon_{ict}$

$\widehat{NRPSReceipt}_{ict}$ is individual receipt of NRPS benefits and we instrument it with $OfferNRPS_{ct}$. $X_{ict}$, is a vector of individual-level controls; $\phi_c$, $\mu_t$ and $\phi_c \times \mu_t$ are community-level, time, and community-time fixed effects, respectively.

*Tobit Model Analysis.* Our final analytical approach tackles a couple of data related issues, both of which have implications for our estimates. The dependent variable in our analysis is self-reported household data on transfers to and from parents. However, since the actual question asked if a transfer (and what amount of transfer) was made in the last twelve months, a couple of data issues could plague our analysis. First, household respondents might mis-specify small transfer amounts from their adult children throughout the year. For instance, it is possible that respondents reported small positive transfers as no transfers (i.e., zero transfers) because respondents forgot the actual amount of money being transferred to them since it is negligible and because it occurred in the past (i.e., right censoring of the dependent variable). Second, and a more important econometric concern, relates to the fact that the dependent variable was captured based on responses elicited for behavior that occurred in the previous twelve months. By the nature of the question (i.e., whether a transfer occurred in the previous twelve months and the actual amount of the transfer), numerous respondents could have actually made transfers to their elderly parents but simply not done so at the time the household survey was administered. This possibility is another example of right-censoring in our outcome variable and one that requires a different approach than the standard OLS estimation. Estimating the effects using OLS or using standard 2SLS will product inconsistent estimates, due to data censoring (Cameron and Trivedi 2005).[47]

---

[45] The results reported in Table 6 provide evidence that there was no statistical difference between the two groups in the period leading up to the implementation of the NRPS.
[46] An assumption of this estimation approach is that the proposed IV variable does not directly influence the outcome.

To tackle the issue of a right-censored dependent variable, we estimate the effects using Tobit-MLE, a method which addresses the two potential sources of right-censoring described above, using the following specification:

$$(4) \quad T^*_{ict} = \beta_0 + \beta_1 (\text{OfferNRPS}_{ct} \times \text{Above60}_{ict}) + \beta_2 \text{Above60}_{ict} + \beta_3 X_{ict} + \phi_c + \mu_t + \phi_c \times \mu_t + \varepsilon_{ict}$$

Where $\varepsilon_{ict} \sim N(0, \sigma^2)$. We observe a latent version of transfers, $T^*_{ict}$, for positive amounts strictly greater than zero. The observed dependent variable is:

$$T_{ict} = \begin{cases} T^*_{ict} & \text{if } T^*_{ict} > 0 \\ 0 & \text{if } T^*_{ict} \leq 0 \end{cases}$$

We compare the marginal effect in (4) to the DDD estimation based on (1). We assume the dependent variable is observed for all positive values of family transfers, and is a linear function of the policy variable and a set of controls. We estimate the Tobit model using a two-way fixed effects model using MLE.

The effect of interest is the marginal effect of the program on actual transfers, not the marginal effect on latent transfers.[48] The marginal effect on latent transfers provides an estimate for uncensored, or desired, transfers. Our objective is to relate the Tobit marginal effects to the DDD estimator in specification (1).[49] Even though the OLS estimate is inconsistent, it should still provide a reasonable approximation of the marginal effect on actual transfers in equation (4). Estimating specification (4) accounts for data censored at zero and enables us to obtain the effect of interest: the marginal effect of the program on the expected value of actual transfers.[50]

---

[47] Several other papers estimate the effect of income (or welfare income) on private transfers, using Tobit analysis alongside OLS for the analysis (Altonji et al. 1996; Raut and Tran 2005; Juarez 2009; Gerardi and Tsai 2014).

[48] The marginal effect on latent transfers is the ML estimated coefficient on $\beta_1$ in (4), $\left( \partial E[T^* | \text{Offer} \times \text{Above60} \, X] / \partial \text{Offer} \times \text{Above60} \right)$.

## VI. Results: NRPS Program Impacts

### A. Results on Program Impacts

*DDD Approach*. In our analysis, we focus on the NRPS impacts on both the extensive (i.e., the impact on the incidence of inter vivos transfers) and intensive margins (i.e., the effect on the amount of inter vivos transfers) of transfers. We begin by estimating the intent-to-treat effects based on specification (1). Table 2, Panel A, reports the OLS estimates. We find that that among NRPS beneficiaries, the access to NRPS benefits leads to a statistically significant reduction of the likelihood that one receives transfers by 6.7 percentage points and lowers the average transfer amount by 9.1 percent (although not significant at the conventional level). The negative estimated coefficients support an altruism model (outlined earlier), whereby an increase in pension income reduces private transfers from adult children. Because the OLS coefficients are likely inconsistent due to right-censoring in the dependent variable, we also report the intent-to-treat effects using a Tobit model.[51]

[Table 2 about here]

Table 2, Panel B, summarizes the Tobit estimates. We focus on the marginal effects: the program effects on transfers for respondents who have positive transfers. The Tobit marginal effects are higher than the OLS (Panel A) as OLS doesn't properly account for observations censored at zero. Using the Tobit estimation, we find that the access to program benefits had an even more pronounced effect on those with positive transfers. In Panel B, we find that among NRPS beneficiaries, access to the NRPS benefits lead to a statistically significant reduction in

---

[49] In equation (2), the coefficient of interest is: $\left(\partial E[T \,|\text{Offer}\times\text{Above60}, \mathbf{X}] / \partial \text{Offer}\times\text{Above60}\right)$

[50] After we estimate the Tobit model in (4), we calculate the marginal effect on actual transfers conditional on them being positive transfers: $\left(\partial E[T \,|\, T>0, \text{Offer}\times\text{Above60}, \mathbf{X}] / \partial \text{Offer}\times\text{Above60}\right) = \beta_1 \Phi(\omega)$, where $\omega = \frac{\beta_1(\text{Offer}\times\text{Above60})}{\sigma}$

[51] We also re-estimated the main OLS specifications but based on the sample used in the Tobit estimations (the Tobit sample is smaller due to issues related of the convergence of the MLE function). Using comparable samples, we find that the pattern and statistical significance results remain consistent with the main results we report.

the probability that one receives transfers from adult children by 7.4 percentage points, and lowered transfer amounts by 14 percent of the average amount.[52] In the DDD approach (Panel A of Tables 2 and Panel A of Table 3), we fail to detect evidence against the null of no statistically significant change in the probability of transfers to children; the Tobit-IV estimation (Panel B of Tables 2 and Panel B of Table 3) bolsters the pattern of the results based on the DDD estimation but is not statistically significant at the conventional levels ($p<0.10$). These estimated negative coefficients support the pattern from recent studies on public pension and private transfers in developing countries (Cox et al. 2004; Juarez 2009; Gerardi and Tsai 2014; Jung et al. 2016) although we report estimates of effect sizes that are much lower than previous empirical estimates. Among the set of studies that use data from other middle-income countries (Juarez 2009, Gerardi and Tsai 2014, Cox, Eser and Jimenez 1998), the estimated likelihood of the pension beneficiary receiving private transfers goes down by approximately 0.48 to 0.55, so, our estimate is significantly smaller than previous estimates from other middle-income countries.[53] Cox et al. (2004) use data from low-income countries but they do not explicitly test for the impact on the likelihood of subsequent intergenerational transfers. In sum, our estimated impacts on subsequent interhousehold transfers are about one-third of previously estimated coefficients from other low-income or middle-income countries.

*2SLS and Tobit Estimates of the Program Impacts.* Table 3 presents these results in Panel A, using the Two-Stage Least Squares (2SLS) approach, and Panel B, using IV-Tobit marginal effects. The instrumental variable approach has very high first stage F-statistics (e.g., the F-statistics associated with the first stage are well above the usual rule of thumb around 10).

[Table 3 about here]

The IV estimates reported in Table 3 provide additional evidence consistent with the pattern of results based on the DDD estimation procedure discussed earlier. As expected, the IV estimate exhibits a slightly higher magnitude of the effect sizes based on the treatment-on-the-treated effects. Panel A shows that NRPS benefit recipients report statistically significant reduction in receiving transfers from children by 9.3 percentage points (baseline mean of 55.1 percent). On the extensive margin, we find that, among program participants, the amount of the transfers received declined by around 13 percent of the average transfer amount. Similar to the result using the DDD estimation, we do not find evidence that the program actually lowered the probability of parents sending income transfers to their children.

---

[52] In addition to the control variables we use reported in the analysis, we also estimated additional analyses where we additionally control for health and income measures, measured at the baseline. The pattern and statistical significance of these additional specifications with additional control variables are consistent with the results we report here.

[53] Using data from South Korea, Jung et al. (2016) estimates a decline of the incidence of transfers by -0.41 percentage points.

Panel B reports the Tobit-IV estimates.[54] Once we account for the right-censoring of the outcome variable and instrument the individual decision to participate in the program, the results corroborate the overall pattern of program impacts (shown in Table 3, Panel B). However, we find a slightly more substantial effect size on the private transfer of beneficiaries. NRPS beneficiaries experienced a statistically significant reduction to receive transfers by 10.4 percentage points (approximately 19 percent). The transfer amount received by benefit recipients declined by 19 percent relative to the average transfer amount received. The effect sizes using the Tobit-MLE procedure are slightly higher than the simple DDD estimated impacts in Table 3. Although the Tobit effect size is somewhat higher than the DDD estimate in Table 3, our estimated coefficients are still considerably lower than previous empirical estimates of the household response from other middle-income countries. Comparing our estimated effect size of the Tobit-MLE procedure to the magnitude of estimates from other middle-income countries based on Juarez (2009), Gerardi and Tsai (2014), and Cox, Eser, and Jimenez (1998), our estimated decline in the likelihood of interhousehold transfers is only one-third of the previously estimated decline in the likelihood of inter vivos transfers. In summary, we find a sizably smaller empirical decline in the likelihood of inter vivos transfers in response to public benefits than previous estimates from other middle-income countries.

Drawing on the results presented in Tables 2 and 3, an important note on how these results relate to the theoretical framework presented earlier is warranted. The pure-altruist model presented in Section III posits a negative relationship between the recipient's publicly funded (i.e., parent's) income and private transfers from the donor (i.e., child). Tables 2 and 3 report results that are consistent with a pure-altruism model, as we detect negative program impacts on both the extensive and intensive margins of private transfers from adult children to their (beneficiary) parent.

---

[54] We used a control function approach to estimate the Tobit-IV based on Rivers and Vuong (1988) and Wooldridge (2010). The approach proceeded in two steps. In the first step we estimated the reduced form regression of endogenous variable on the instrument and controls using OLS; in the second step, we estimated a Tobit model using MLE of the dependent variable on the endogenous variable, the OLS residual from first and controls; we clustered standard errors are clustered at the community level; then, we calculated the marginal effects.

## B. The NRPS Crowd-Out Effect

Next, we investigate whether changes to individual income induced by participation in the NRPS led to changes in total inter vivos transfers to the program beneficiary (i.e., the crowd-out effect). As highlighted earlier in the conceptual framework, when an individual receives NRPS benefits, his/her income level increases; a change that could induce a large equilibrium response in private transfers depending on household preferences. In fact, our model suggests a quantitatively large crowd-out effect induced by the public program. We estimate the crowd-out effect using a 2SLS approach (similar to specification 4). We estimate the crowd-out effect using the following specification[55]:

$$(5) \quad T_{ict} = \beta_0 + \beta_1 \widehat{Benefits}_{ict} + \beta_2 X_{ict} + \phi_c + \mu_t + \varepsilon_{ict}$$

Where, $T_{ict}$ is the transfer variable of interest, $X_{ict}$ is a vector of control variables, and $\beta_1$ is the crowd-out effect. The coefficient $\beta_1$ is the yuan for yuan crowd-out, and not a measure of elasticity. We instrument $Benefits_{ict}$ (NRPS benefit amount in yuan) with $OfferNRPS_{ct}$ (an indicator equaling 1 if the community was offered access to NRPS at time $t$). Both transfers and NRPS benefits are in yuan terms, not in log terms. Therefore, we estimate a crowd-out measure and not an elasticity. [56]

[Table 4 about here]

Table 4 reports the results on the estimated crowd-out effects in two panels (based on two different estimation approaches). Panel A reports the estimates based on the reduced-form approach without accounting for the right-censoring of the outcome variable. Alternatively, Panel B reports the Tobit-IV estimates; this estimate is the one we focus on. Panel B reports a small, negative but imprecisely estimated effect size. In comparison to estimates of the crowd-out effect found in other middle-income studies, this study's estimated crowd-out effect is significantly lower. For example, using data from Mexico, Juarez (2009) estimates a crowd-out effect of -0.86. Jensen (2004) estimates an average crowd-out effect of -0.3 using data from

---

[55] This approach closely follows the approach for estimating the crowd-out effect in Juarez (2009) and Gerardi and Tsai (2014), who also study interhousehold transfers in response to public programs (in Mexico and Taiwan respectively).

[56] Some empirical papers estimate a log-log specification and then multiply their elasticity measures by a ratio of Y to X to get a crowd-out effect, or the partial derivative of transfers. Their crowd-out effects vary by choice of mean, median or quantile values for the ratio of Y to X. As noted in Payne (1998), the model specification matters. In a log-log linear specification, $\beta_1$ is the elasticity being estimated. The main assumption here is that the elasticity is constant, which is the opposite of the assumption made in Section 3--that the transfers-benefits gradient is -1. Following previous literature, we do not use a log-log specification.

South Africa. The absolute value of the estimated coefficient based on the Tobit-IV estimation (i.e., crowd-out effect of -0.03) is the lowest in magnitude of found among estimates from other middle-income countries although the effect size is imprecisely estimated.[57]

C. **Heterogeneity of Program Impacts by Income Level**

The amount of transfers could vary by the level of income of the recipient. Therefore, we examine for program impact across two levels of income levels, measured at baseline in the CHARLS survey. Specifically, we focus on the magnitude of the crowd-out effect by the households' income level (i.e., treatment heterogeneity). Treatment heterogeneity could also have important policy implications that can inform better program targeting. To this end, we estimate the local average treatment effects by income level and we repeat specifications (3) and (4) for two samples: for households below and for households above the poverty-level income threshold.[58]

[Table 5a about here]  [Table 5b about here]

We report the results from the two specifications in Table 5a and Table 5b. Both tables, Table 5a and Table 5b, report the estimates for the crowd-out effect. However, each table reports results based on different definitions of the independent variable. In Table 5a, we employ a definition of the independent variable that is based on the amount of yuan received as an independent variable. In contrast, Table 5b reports the results based on a specification in which the independent variable is defined as a binary variable (e.g., whether a public benefit was reported, or not). Although the two tables use alternative definitions for the independent variable, both tables report crowd-out effect estimates in response to these alternative definitions of the independent variable. First, we focus our attention on the estimated marginal effects ($E[T \mid T>0]$) based on the Tobit-IV estimation, reported in Panel B of Table 5a. As highlighted earlier, this particular estimation address the possibility that unobservable characteristics at the community level could bias estimates of the estimated coefficient for the crowd-out effect and the right-censoring in our outcome variable. The estimated coefficient, based on the two-stage least-squares Tobit estimation (Panel A), is -0.37 for households who report below the poverty line income level (reported in Table 5a, Column 1) and 0.04 for households living above the poverty line (reported in Table 5a, Column 2). These estimated effect sizes suggest that the estimated crowd-out effect is stronger (albeit we caveat that these estimates are imprecisely estimated) for households living below the poverty line. Although this finding is consistent with previous estimates, the magnitude of the crowd-out effect is much smaller as compared to the estimated effect sizes from data and interventions from middle-income countries. For example, Cox et al. (2004) examine crowd-out effects under an Indonesian pension program and find that the level of crowd-out is more severe for low-income households.[59]

---

[57] Only one other study, Cox, Hansen and Jimenez (2004), finds a similar estimate to ours, but the study examines the response using data from a low-income country (Philippines).
[58] We use the World Bank individual income classification of poverty based on the 1.90 USD/ daily threshold. Based on the World Bank's Millennium Development Goals, China classifies the poverty standard as having a per capita income of 2,300 in 2010 yuan (World Bank, 2017). The standard is approximately equal to $1.90 per day in purchasing parity terms using the 2010 yuan to U.S. conversion. In 2010, $1 was worth 3.31 yuan in purchasing parity terms. This yields a conversion of 2296 yuan, annually.

Table 5b reports results for the heterogeneous treatment effects (by income level) for the NRPS program participation but the reported results are based on specifications relying on a binary definition of the independent variable (i.e., receipt of NRPS benefits, or not). We estimated specifications (3), and (4) for respondents living at/below the poverty line and for those living above the poverty line. When we focus our attention on the Tobit-IV estimated effect size, and consistent with the results from Table 5a, we find suggestive evidence that the likelihood of program beneficiaries receiving no inter vivos transfers decreases among households reporting baseline income below the poverty level (as reported in Table 5b, Column 1 for the marginal effect). When we turn our attention to Tobit-IV estimates for the sample reporting baseline income above the poverty level, the estimated effect size on the crowd-out effect was larger than the estimated crowd-out effect for the below the poverty line sample (i.e., the marginal effect reported in Table 5b, Column 3). However, the estimated effect size is very statistically significant.[60,61] These estimates suggest stronger behavioral response to the NRPS program for high-income households than for low-income households. From a program targeting perspective, this implies that the policy may be welfare-enhancing for low-income beneficiaries.

---

[59] Cai et al. (2006) find similar evidence in urban China for retired urban pensioners. They study the question from a different angle and find that the responsiveness of transfers falls as income increases. Suggesting a larger crowd-out effect for low-income households in urban China.
[60] For beneficiaries living below the poverty line, we still detect evidence that they receive fewer inter vivos transfers but the estimated effect size is insignificant.
[61] We estimated these regressions using alternative income thresholds and the results remain consistent. Tables using alternative income-status thresholds are reported in Online Appendix B.

# VII. Robustness Checks of Program Impacts

## A. Falsification Tests

To test for potential spurious effects, we also perform a falsification exercise based on specifications (1) and (4) but we draw on data from an alternative group of individuals. Specifically, we draw on an alternative "placebo" sample of group of individuals that, in theory, should not be affected by the program. The NRPS program is available to individuals living in rural administrative districts who aren't already enrolled in an urban pension scheme. Additionally, one feature of the NRPS program is that age-eligible individuals can buy their way into the program if their children contribute on their behalf. In the main analysis, we used survey responses to remove urban pensioners and elderly individuals without children that live in rural administrative districts (rural *Hukou*). However, for this falsification exercise, we reconstruct a sample based only including these respondents. In the falsification exercise, we only include (1) individuals who live in rural areas but are obtaining benefits from an urban pension system, or (2) individuals who are elderly and without children who happen to live in rural administrative districts and have no pension benefits from any other program (urban or rural). The objective of this falsification exercise is to examine whether the transfer outcomes for this "placebo" sample would differ between areas that had NRPS coverage and areas that did not: the placebo sample is not directly affected by the NRPS program in theory, so this falsification exercise should yield non-significant results for NRPS program impacts on the transfer outcomes.

Using this alternative sample, we then re-estimate specifications (1) and (4), as outlined in the main analysis, but the estimation is based only on this alternative sample. Appendix Table A.2 reports the results. Panel A (Columns 1 through 4) reports the results for the OLS estimation; Panel B reports the estimated program impacts among the alternative sample (Columns 1 through 4 report the marginal effects, $E[T \mid T>0]$). Appendix Table A.2 reports estimates of $\beta_1$ (from the OLS and Tobit specifications highlighted earlier). The estimates of the effect size are not-significant for any of the study outcomes, as theoretically expected. Therefore, the falsification exercise further bolsters the validity of our DDD empirical design.[62]

---

[62] The estimations for the Tobit-IV estimation for the transfers from children outcome is based on 746 urban pensioners so we caveat that our results for this falsification exercise are the best available results given the data available and we acknowledge the possibility of the estimations being statistically underpowered.

### B. Placebo Outcome Tests

As an additional check for the plausibility of the results based on our main specification, we also re-estimate (1) and (4) on a set of outcomes for which no theoretical basis exists to detect program impacts (i.e., placebo outcomes). For this particular exercise, we choose outcomes with no plausible mechanism by which these "placebo outcomes" could have been caused by changes in NRPS program participation. Therefore, any significant correlation would be from random chance or an artifact of the research design that might also have given rise to our main results highlighted earlier. Specifically, we perform this additional falsification exercise with placebo outcomes: the likelihood of one's nationality being Han and the number of daughters in the household. In Appendix Table A.3, we report the DDD results (for the ITT and TOT specifications) on the two placebo outcomes: the likelihood of one's nationality being Han and the number of daughters in the household. Panel A reports the intent-to-treat results and Panel B reports the treatment-on-treated results. Given the high number of observations in the sample for these two outcomes, the reported results fail to detect evidence of any program impacts on the two placebo outcomes. Since these outcomes were not supposed to be affected by the NRPS reform, this additional falsification exercise further bolsters the validity of our estimation approach.

## VIII. Discussion and Conclusion

China is facing rapid population aging, a phenomenon that strained the country's defined benefit pension system and the country's traditional family-based old-age support system in 2000s. In 2009, three out of four Chinese workers had no pension coverage. Furthermore, demographic projections show that by 2030, China will be home for approximately a quarter of the world's population aged 60 and above. The number of Chinese individuals aged sixty and above will be larger than that of all the European countries combined: it will be equal to the elderly population of North America, Latin America and the Caribbean, Africa, and Oceania (UN 2010). Given the size of its aging population, the Chinese government gradually took on the responsibility to alleviate old-age poverty via pension reforms. Given the size of the country's population, the implications of what happens to the pension reform of China's old-age pension system can have tremendous consequences.

We explore the inter vivos behavioral response to China's new rural pension scheme. Specifically, we examine whether receiving pension benefits induces a reduction in private transfers from adult children and estimate the crowd-out effect of public pension benefits to the elderly. Previous studies that examine inter vivos transfers in developing and developed countries find a negative relationship and substantially large substitution between private transfers and the recipient's income. In this paper, we provide early estimates on how pension benefits influence inter vivos decisions in the context of a developing country. We estimate the impact of a public pension program in the context of the largest country in the world, with a population that is rapidly ageing. Using a DDD strategy coupled with a Tobit-2SLS estimation, we find that the receipt of NRPS benefits reduces the likelihood of NRPS beneficiaries receiving private transfers from their children, even though the magnitude of reduction is quantitatively small. Our findings are consistent with previous empirical studies on this relationship, but we estimate a substantially smaller effect size (than previous studies in the context of developing countries) of the private inter vivos transfer response from the children to the parents in response the public program.

We also empirically estimate for the magnitude of the crowd-out effect among eligible individuals receiving benefits and we find some evidence of the crowding-out of private household transfers, but the estimated effect size is close to zero and statistically insignificant. In comparison to previous studies from middle-income countries, our estimates on the NRPS program impact on both the incidence of inter vivos transfers and on magnitude of the crowd-out effect are notably smaller. Overall, our estimates of the crowd-out effect and in particular the smaller effect size in comparison with previous studies in developing countries could have several explanations.  One interpretation for the estimated effect size could be due to the fact that family ties are strong in China given the tradition of filial piety and close ties between parents and children. Another potential interpretation is that the increase in NRPS benefits are smaller in rural China. Therefore, the modest pension benefits in rural China may be unlikely to crowd out the inter vivos motives of adult children for poverty alleviation of their parents. A third potential explanation could be that the NRPS program is relatively recent. It could be that the salience of the NRPS benefits hasn't quite been fully taken into consideration by redistributing transfers among members of the same household.

These small program impacts on subsequent inter vivos transfers from household members imply that the introduction of pecuniary pension benefits can have additional welfare-improving consequences, contrary to the results based on previous empirical studies from other middle-income countries. Our findings have important welfare implications for the poorest households, in particular. One of the objectives of the NRPS program was to ensure the well-being of the aging population and to mitigate the risk of old-age poverty in rural areas, where the poverty rate is already very high. Although conceptually the welfare effects can be ambiguous *ex ante*, any empirical behavioral response and tendency to further crowd out familiar transfers to the elderly may be of high concern in a program for which one of the objectives is to address old-age poverty. Yet, among the poorest households in our sample, we find no evidence to suggest that there is significant crowding out of familiar transfers to elderly parents as a result of increased NRPS benefits and the estimated effect size in that sub-sample is smaller than the one based on the whole sample. Thus, among this demographic, any *ex ante* concerns for disincentives for family transfers appear second order although understanding the mechanisms behind this interplay between the poverty status and the strength of the inter vivos transfers' response merits further research.

The findings reported in this paper are important because we show that a pension benefit reform in the largest country could produce pecuniary benefits for program beneficiaries without spurring the negative behavioral response from household members, as suggested by studies in high-income or middle-income countries. This study concentrated on estimating the inter vivos transfers to program beneficiaries. The NRPS intervention was introduced in part to mitigate the risk of old-age poverty. This large public program, however, may have also had additional broader impacts on the Chinese population, its economy and on income inequality. How the NRPS program influenced these outcomes should be the subject of future work.

# Figures and Tables

(a) 2009-2010  (b) 2010-2013

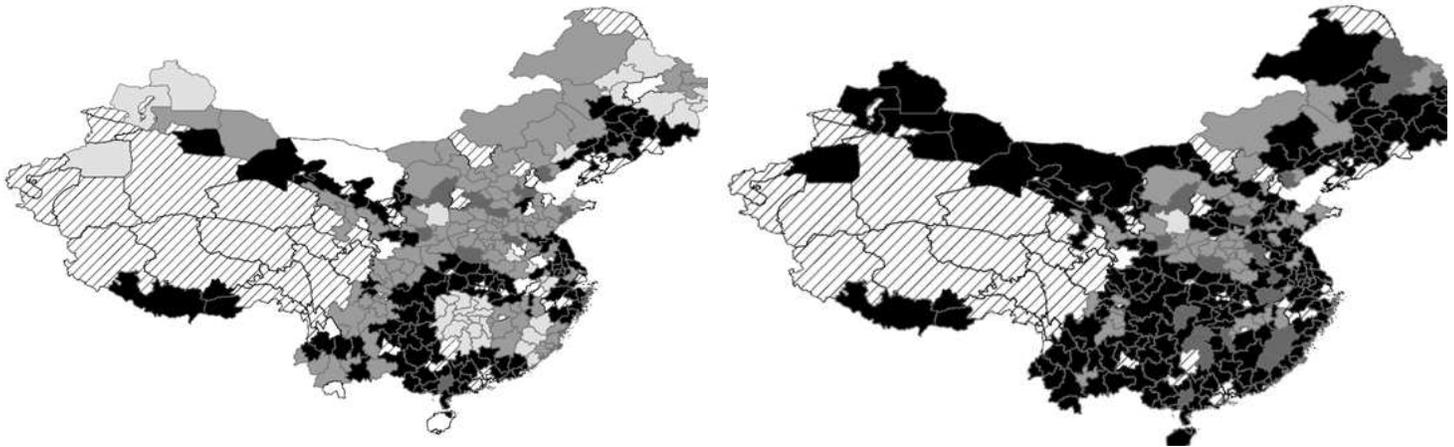

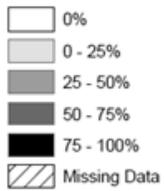

**Fig 1.** Geographic Implementation of NRPS.

Participating Provinces in China Health and Nutrition Survey

Sampled Communities in the China Health and Retirement Longitudinal Study

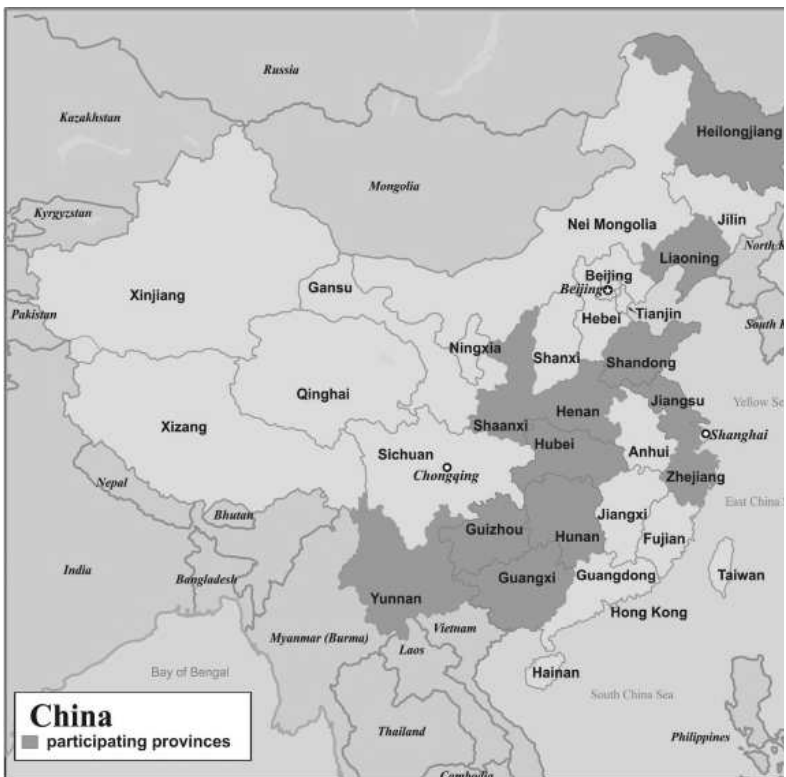
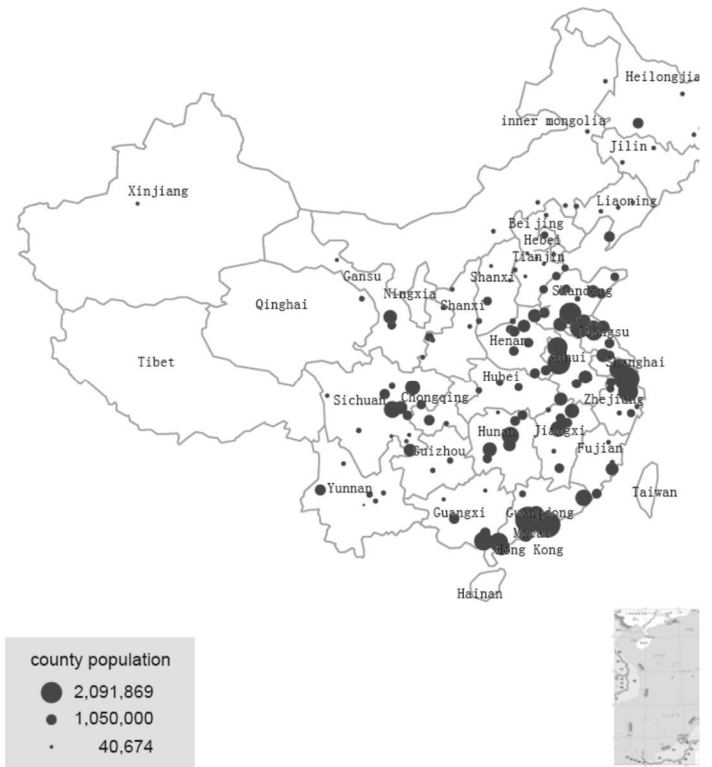

**Fig 2.** Coverage Maps.

**Table 1:** Summary Statistics.

| | Full Sample | Baseline Participants | Baseline Non-Participants | p-value[a] |
|---|---|---|---|---|
| *Demographics of Respondents* | | | | |
| Respondent's Age | 59.31 (10.01) | 58.43 (9.68) | 58.44 (10.24) | 0.99 |
| # of Household Residents | 3.74 (1.87) | 3.68 (1.78) | 3.75 (1.88) | 0.04 |
| Percent Female | 0.53 (0.50) | 0.54 (0.50) | 0.53 (0.50) | 0.38 |
| Percent Married | 0.80 (0.40) | 0.81 (0.39) | 0.78 (0.41) | 0.00 |
| Percent With At Least Lower Secondary Education | 0.48 (0.50) | 0.48 (0.50) | 0.46 (0.50) | 0.10 |
| Weekly Work Hours | 45.45 (23.87) | 47.26 (24.07) | 46.89 (22.70) | 0.50 |
| Percent Currently Working | 0.70 (0.46) | 0.70 (0.46) | 0.69 (0.46) | 0.11 |
| Percent Working in Agriculture | 0.72 (0.45) | 0.72 (0.45) | 0.73 (0.45) | 0.49 |
| | | | | |
| *Health Domains and Health Status* | | | | |
| Percent with Difficulty Climbing Flight of Stairs | 0.43 (0.49) | 0.40 (0.49) | 0.43 (0.50) | 0.01 |
| Mobility Index [b] | 0.00 (1.37) | -0.06 (1.35) | 0.00 (1.37) | 0.02 |
| Percent Rarely Feeling Happy | 0.25 (0.43) | 0.17 (0.38) | 0.19 (0.39) | 0.08 |
| Affect Index [b] | 0.00 (1.19) | -0.04 (1.17) | -0.01 (1.24) | 0.27 |
| Short Distance Vision is At Least Good | 0.33 (0.47) | 0.33 (0.47) | 0.33 (0.47) | 0.80 |
| Vision Index [c] | 0.00 (1.19) | -0.03 (1.19) | 0.03 (1.21) | 0.02 |
| Percent with Difficulty Taking Bath | 0.07 (0.26) | 0.06 (0.23) | 0.08 (0.27) | 0.00 |
| Self-Care Index [b] | 0.00 (1.43) | -0.07 (1.34) | 0.06 (1.55) | 0.00 |
| Percent with Difficulty Cleaning House | 0.11 (0.31) | 0.09 (0.29) | 0.11 (0.31) | 0.00 |
| Usual Activities Index [b] | 0.00 (1.47) | -0.06 (1.38) | 0.03 (1.51) | 0.00 |
| Percent with Weekly Contact with Parents or In-Laws | 0.27 (0.44) | 0.30 (0.46) | 0.28 (0.45) | 0.13 |
| Social Interactions Index [c] | 0.00 (1.19) | 0.15 (1.12) | 0.01 (1.16) | 0.00 |
| Percent with At Least Good Self-Reported Health Status | 0.25 (0.43) | 0.27 (0.44) | 0.26 (0.44) | 0.23 |
| | | | | |
| *Health Behavior, Healthcare Utilization and Disease States* | | | | |
| Percent Ever Smoked | 0.41 (0.49) | 0.40 (0.49) | 0.40 (0.49) | 0.98 |
| Percent Smoking Now | 0.25 (0.44) | 0.29 (0.45) | 0.30 (0.46) | 0.39 |
| Percent Consuming Alcohol in Past Year | 0.33 (0.47) | 0.33 (0.47) | 0.33 (0.47) | 0.74 |
| Percent Drinking At Least Once a Week | 0.18 (0.39) | 0.16 (0.37) | 0.16 (0.36) | 0.28 |
| Sleep Duration (hours per night) | 6.28 (1.94) | 6.40 (1.93) | 6.32 (2.00) | 0.07 |
| Number of Doctor Visits | 0.20 (0.40) | 0.20 (0.40) | 0.19 (0.39) | 0.07 |
| Number of Nights Stayed at the Hospital | 0.11 (0.31) | 0.10 (0.29) | 0.09 (0.28) | 0.06 |
| Percent Ever Diagnosed with Diabetes | 0.05 (0.22) | 0.05 (0.22) | 0.04 (0.20) | 0.01 |
| Percent Ever Diagnosed with Hypertension | 0.25 (0.43) | 0.25 (0.43) | 0.22 (0.42) | 0.01 |
| | | | | |
| Observations | 28,034 | 10,011 | 3,680 | |

Notes: Standard deviations are reported in parenthesis. (a) We test the null hypothesis that the difference in participant and non-participant means is equal to 0. (b) Measured in terms of difficulty. Low (or Negative) values denote less difficulty and better mood/emotion. (c) Positively coded where Higher values denote better vision and increased social interaction.

**Table 2:** Intent-to-Treat (ITT) Estimates on Transfer Outcomes (DDD).

|  | Transfer From Children (Yes=1) | Transfer To Children (Yes=1) | Transfer From Children (Yuan) | Transfer To Children (Yuan) |
|---|---|---|---|---|
|  | (1) | (2) | (3) | (4) |
| Panel A (OLS): |  |  |  |  |
| Offered NRPS * Above60[a] | -0.067*** | -0.047* | -385.925 | -1.977 |
|  | (0.017) | (0.021) | (322.244) | (263.450) |
| Baseline Mean | 0.551 | 0.161 | 4242.603 | 2847.255 |
| Controls | Yes | Yes | Yes | Yes |
| R-squared | 0.340 | 0.193 | 0.102 | 0.083 |
| Observations | 11,562 | 11,540 | 11,537 | 11,534 |
| Panel B (TOBIT)[b]: |  |  |  |  |
| Marginal effect, E(T) | -0.129*** | -0.133** | -1639.206*** | -1761.077* |
|  | (0.028) | (0.066) | (538.153) | (1025.290) |
| Marginal effect, E[T \| T>0] | -0.074*** | -0.040** | -576.194*** | -426.139* |
|  | (0.016) | (0.020) | (187.4478) | (245.3874) |
| Baseline Mean | 0.551 | 0.161 | 4242.603 | 2847.255 |
| Controls | Yes | Yes | Yes | Yes |
| Pseudo R-squared | 0.173 | 0.094 | 0.013 | 0.017 |
| Observations | 9,210 | 8,231 | 9,173 | 8,225 |

Notes: (a) Policy instrument interacted with an indicator for being over 60 years old. (b) Tobit panel: Top row is the partial effect on latent dependent variable. Bottom row is the marginal effect on positive transfers (the effect of interest). Individual level controls: Age, Age Squared, Marital Status (=1 if Married), Gender (=1 if Female), Education Levels (Base Group is illiterate with no formal education), Household Size (# of Household Residents). All regressions are estimated with Community, Year and (Community X Year) fixed effects. Clustered standard errors at the community level reported in parenthesis. *p< 0.10, **p< 0.05, ***p< 0.01.

**Table 3:** Treatment-on-treated (TOT) Estimates on Transfer Outcomes (DDD).[a]

|  | Transfer From Children (Yes=1) | Transfer To Children (Yes=1) | Transfer From Children (Yuan) | Transfer To Children (Yuan) |
|---|---|---|---|---|
|  | (1) | (2) | (3) | (4) |
| Panel A (2SLS): |  |  |  |  |
| NRPS Receipt* Above60[b] | -0.093*** | -0.065** | -535.337 | -2.748 |
|  | (0.023) | (0.029) | (447.702) | (366.131) |
| Baseline Mean | 0.551 | 0.161 | 4242.603 | 2847.255 |
| Controls | Yes | Yes | Yes | Yes |
| F-Stat (First Stage) | 537.74 | 535.35 | 536.23 | 534.87 |
| R-squared | 0.341 | 0.193 | 0.102 | 0.083 |
| Observations | 11,562 | 11,540 | 11,537 | 11,534 |
| Panel B (TOBIT-IV)[c]: |  |  |  |  |
| Marginal effect, E(T) | -0.180*** | -0.183** | -2276.892*** | -2429.474* |
|  | (0.028) | (0.090) | (745.359) | (1411.313) |
| Marginal effect, E[T \| T>0] | -0.104*** | -0.055** | -799.732*** | -590.058* |
|  | (0.022) | (0.027) | (259.123) | (340.079) |
| Baseline Mean | 0.551 | 0.161 | 4242.603 | 2847.255 |
| Controls | Yes | Yes | Yes | Yes |
| F-Stat (First Stage) | 480.38 | 461.90 | 479.98 | 461.90 |
| Pseudo R-squared | 0.174 | 0.095 | 0.013 | 0.017 |
| Observations | 9,210 | 8,231 | 9,173 | 8,225 |

Notes: (a) The Treatment on the Treated (TOT) effects. (b) Endogenous NRPS Receipt is instrumented by a policy variable (Offer NRPS). (c) Tobit-IV panel: Top row is the partial effect on latent dependent variable. Bottom row is the marginal effect on positive transfers (the effect of interest). Individual level controls: Age, Age Squared, Marital Status (=1 if Married), Gender (=1 if Female), Education Levels (Base Group is illiterate with no formal education), Household Size (# of Household Residents). All regressions are estimated with Community, Year and (Community X Year) FE. Clustered standard errors at the community level reported in parenthesis. *p< 0.10, **p< 0.05, ***p< 0.01.

Table 4: Crowd-Out Effects Using Main Specification.[a]

| | Transfer From Children (Yuan) | Transfer To Children (Yuan) |
|---|---|---|
| | (1) | (2) |
| Panel A (2SLS): | | |
| NRPS Benefit (Yuan) * Above60 [b] | -0.623 | -0.524 |
| | (0.512) | (0.461) |
| Baseline Mean | 4242.603 | 2847.255 |
| Controls | Yes | Yes |
| F-Stat (First Stage) | 34.49 | 34.70 |
| R-squared | 0.049 | 0.023 |
| Observations | 11,696 | 11,740 |
| Panel B (TOBIT-IV)[c]: | | |
| Marginal effect, E(T) | -0.084 | 1.010 |
| | (0.243) | (0.798) |
| Marginal effect, E[T \| T>0] | -0.030 | 0.238 |
| | (0.086) | (0.188) |
| Baseline Mean | 4242.603 | 2847.255 |
| Controls | Yes | Yes |
| F-Stat (First Stage) | 45.24 | 45.14 |
| Pseudo R-squared | 0.014 | 0.016 |
| Observations | 11,717 | 11,760 |

Notes: (a) The crowd-out effect is estimated by regressing transfer amount on NRPS benefit amount interacted with age indicator. (b) Endogenous NRPS Benefit Amount instrumented by policy instrument (NRPS Offer). (c) Tobit- IV panel: Top row is the partial effect on latent dependent variable. Bottom row is the marginal effect on positive transfers (the effect of interest). Individual level controls: Age, Age Squared, Marital Status (=1 if Married), Gender (=1 if Female), Education Levels (Base Group is illiterate with no formal education, Household Size (# of Household Residents). All regressions are estimated with Community, Year and (Community X Year) fixed effects. Clustered standard errors at the community level reported in parenthesis. *p< 0.10, **p< 0.05, ***p< 0.01.

**Table 5a:** Heterogeneous Crowd-Out Effect by Pre-Transfer Income.[a]

| | Transfer From Children (Yuan) [b] | | Transfer To Children (Yuan) [c] | |
|---|---|---|---|---|
| | Pre-Transfer Income <= Poverty Line | Pre-Transfer Income > Poverty Line | Pre-Transfer Income <= Poverty Line | Pre-Transfer Income > Poverty Line |
| | (1) | (2) | (3) | (4) |
| Panel A (2SLS): | | | | |
| NRPS Receipt * Above60 [d] | 0.410 | -1.271* | -1.235** | -0.510 |
| | (1.764) | (0.743) | (0.616) | (0.797) |
| Baseline Mean | 2372.711 | 2372.711 | 1425.112 | 1425.112 |
| Controls | Yes | Yes | Yes | Yes |
| F-Stat (First Stage) | 239.29 | 51.64 | 247.01 | 47.87 |
| R-squared | 0.159 | 0.058 | 0.080 | 0.030 |
| Observations | 3,467 | 5,552 | 3,483 | 5,569 |
| Panel B (TOBIT-IV) [e]: | | | | |
| Marginal effect, E(T) | -0.986 | 0.105 | -0.126 | 1.175 |
| | (1.357) | (0.222) | (1.650) | (1.158) |
| Marginal effect, E[T \| T>0] | -0.374 | 0.038 | -0.028 | 0.289 |
| | (0.506) | (0.079) | (0.365) | (0.285) |
| Baseline Mean | 2372.711 | 2372.711 | 1425.112 | 1425.112 |
| Controls | Yes | Yes | Yes | Yes |
| Pseudo R-squared | 0.021 | 0.020 | 0.039 | 0.022 |
| Observations | 3,486 | 5,577 | 3,502 | 5,595 |

Notes: (a) Based on the World Bank's Millennium Development Goals, China classifies the poverty standard as per capita income of 2,300 in 2010 yuan (World Bank, 2017). (b) Transfer amount from children (reported in yuan). (c) Transfer amount to children (reported in yuan). (d) Endogenous NRPS benefit amount is instrumented by a policy variable (Offer NRPS). (e) Tobit-IV panel: Top row is the partial effect on latent dependent variable. Bottom row is the marginal effect on positive transfers (the effect of interest). Individual level controls: Age, Age Squared, Marital Status (=1 if Married), Gender (=1 if Female), Education Levels (Base Group is illiterate with no formal education), Household Size (# of Household Residents). All regressions are estimated with Community, Year and (Community X Year) fixed effects. Clustered standard errors at the community level reported in parenthesis. *p< 0.10, **p< 0.05, ***p< 0.01.

**Table 5b:** Heterogeneous Crowd-Out Effect by Pre-Transfer Income.[a]

| | Pre-Transfer Income <= Poverty Line | | Pre-Transfer Income > Poverty Line | |
|---|---|---|---|---|
| | Transfer From Children (Yes=1) [b] | Transfer From Children (Yuan) [c] | Transfer From Children (Yes=1) [b] | Transfer From Children (Yuan) [c] |
| | (1) | (2) | (3) | (4) |
| Panel A (2SLS): | | | | |
| NRPS Receipt* Above60 [d] | -0.017 | 95.701 | -0.101*** | -955.016 |
| | (0.049) | (1242.683) | (0.036) | (660.895) |
| Baseline Mean | 0.475 | 2160.544 | 0.333 | 1480.863 |
| Controls | Yes | Yes | Yes | Yes |
| F-Stat (First Stage) | 653.74 | 640.80 | 519.93 | 519.38 |
| R-squared | 0.248 | 0.083 | 0.253 | 0.060 |
| Observations | 2,956 | 2,948 | 5,208 | 5,200 |
| Panel B (TOBIT-IV) [e]: | | | | |
| Marginal effect, E(T) | -0.047 | -225.859 | -0.208*** | -2797.418*** |
| | (0.077) | (1727.011) | (0.057) | -934.572 |
| Marginal effect, E[T \| T>0] | -0.029 | -83.485 | -0.112*** | -973.214*** |
| | (0.048) | (635.447) | (0.031) | (323.632) |
| Baseline Mean | 0.475 | 2160.544 | 0.333 | 1480.863 |
| Controls | Yes | Yes | Yes | Yes |
| F-Stat (First Stage) | 808.47 | 808.47 | 619.69 | 619.69 |
| Pseudo R-squared | 0.183 | 0.016 | 0.169 | 0.016 |
| Observations | 2,956 | 2,948 | 5,208 | 5,200 |

Notes: (a) Based on the World Bank's Millennium Development Goals, China classifies the poverty standard as per capita income of 2,300 in 2010 yuan (World Bank, 2017). (b) Transfer amount from children (=1 if Yes). (c) Transfer amount to children (reported in yuan). (d) Endogenous NRPS benefit receipt (=1 if Began Receiving Benefits) is instrumented by a policy variable (Offer NRPS). (e) Tobit-IV panel: Top row is the partial effect on latent dependent variable. Bottom row is the marginal effect on positive transfers (the effect of interest). Individual level controls: Age, Age Squared, Marital Status (=1 if Married), Gender (=1 if Female), Education Levels (Base Group is illiterate with no formal education), Household Size (# of Household Residents). All regressions are estimated with Community, Year and (Community X Year) fixed effects. Clustered standard errors at the community level reported in parenthesis. *p< 0.10, **p< 0.05, ***p< 0.01.

# Online Appendix
## A. Robustness Checks

Table A.1: Test of Common Trends Using CHNS Data.

|  | Transfer From Children (Yuan) | In-Kind Transfer From Children (Yes=1) | In-Kind Transfer From Children (Yuan) |
|---|---|---|---|
|  | (1) | (2) | (3) |
| Treatment * 2004 | 697.621 | -0.042 | -83.105 |
|  | (692.830) | (0.086) | (104.600) |
| Treatment * 2009 | 806.916 | -0.112 | -29.249 |
|  | (645.392) | (0.085) | (106.388) |
| R-Squared Adj | 0.161 | 0.074 | 0.151 |
| Year FE | Yes | Yes | Yes |
| Community FE | Yes | Yes | Yes |
| Observations | 7,032 | 18,465 | 9,544 |

Notes: Base year is 2006. Columns 1-3 are estimated using Ordinary Least Squares (OLS) with Community and Year FE. Clustered standard errors at the community level reported in parenthesis. *$p< 0.10$, **$p< 0.05$, ***$p< 0.01$. Source: CHNS 2004, 2006 and 2009 waves.

Table A.2: Falsification Test Using Placebo Sample[a].

| Panel A (OLS): | Transfer From Children (Yes=1) | Transfer To Children (Yes=1) | Transfer From Children (Yuan) | Transfer To Children (Yuan) |
|---|---|---|---|---|
| | (1) | (2) | (3) | (4) |
| Offered NRPS * Above60 [b] | 0.051 | 0.009 | 127.572 | 1533.332 |
| | (0.074) | (0.082) | (1330.410) | (1815.857) |
| Baseline Mean | 0.233 | 0.214 | 7,576.844 | 16,503.690 |
| Controls | Yes | Yes | Yes | Yes |
| R-squared | 0.553 | 0.482 | 0.297 | 0.357 |
| Observations | 685 | 685 | 685 | 685 |

| Panel B (TOBIT) [c]: | | | | |
|---|---|---|---|---|
| | (1) | (2) | (3) | (4) |
| Marginal effect, $E(T)$ | -0.210 | 0.070 | -5075.081 | -2252.211 |
| | (0.163) | (0.259) | (3577.173) | (6663.510) |
| Marginal effect, $E[T \mid T>0]$ | -0.053 | 0.014 | -944.095 | 372.087 |
| | (0.040) | (0.053) | (634.858) | (1111.883) |
| Baseline Mean | 0.233 | 0.214 | 7,576.844 | 16,503.690 |
| Controls | Yes | Yes | Yes | Yes |
| R-squared | 0.445 | 0.365 | 0.086 | 0.070 |
| Observations | 875 | 875 | 875 | 875 |

Notes: (a) The placebo sample comprises two groups that in theory should not be affected by the NRPS program: the first group is pensioners who live in rural areas but receive benefits from an urban pension program; the second group is elderly individuals who are ages 60 and above and without any formal pension benefits. (b) Our DDD coefficient: Policy instrument interacted with an indicator for being over 60 years old. A significant coefficient would imply the presence of treatment effects for the "placebo" group in treated communities. (c) Tobit panel: Top row is the partial effect on latent dependent variable. Bottom row is the marginal effect on positive transfers (the effect of interest). Individual level controls: Age, Age Squared, Marital Status (=1 if Married), Gender (=1 if Female), Education Levels (Base Group is illiterate with no formal education), Household Size (# of Household Residents). All regressions are estimated with Community, Year and (Community X Year) FE. Clustered standard errors at the community level reported in parenthesis. *p< 0.10, **p< 0.05, ***p< 0.01.

**Table A.3:** Falsification Test Using Placebo Outcomes.

|  | Nationality (Han =1) | Daughters (#) |
|---|---|---|
|  | (1) | (2) |
| Panel A (ITT): |  |  |
| Offered NRPS * Above60 [a] | 0.013 | 0.006 |
|  | (0.008) | (0.044) |
| Baseline Mean | 0.920 | 1.299 |
| Controls | Yes | Yes |
| R-squared | 0.636 | 0.147 |
| Observations | 12,107 | 12,107 |
| Panel B (TOT): |  |  |
| NRPS Receipt * Above60 [b] | 0.018 | 0.082 |
|  | (0.011) | (0.060) |
| Baseline Mean | 0.920 | 1.299 |
| Controls | Yes | Yes |
| F-Stat (First Stage) | 797.19 | 797.19 |
| R-squared | 0.636 | 0.147 |
| Observations | 12,107 | 12,107 |

Notes: (a) Our ITT coefficient: Policy instrument interacted with an indicator for being over 60 years old. The control group becomes individuals under the Age of 60 living in eligible communities that didn't offer NRPS between '11 and '13. (b) Individual participation instrumented with the policy variable. Individual level controls: Age, Age Squared, Marital Status (=1 if Married), Gender (=1 if Female), Education Levels (Base Group is illiterate with no formal education), Household Size (# of Household Residents). Panel A is estimated using Ordinary Least Squares (OLS) with Community, Year and Community*Year FE. Panel B is estimated using Two-Stage Least Squares (2SLS) with Community, Year and Community*Year FE. Clustered standard errors at the community level reported in parenthesis. *$p< 0.10$, **$p< 0.05$, ***$p< 0.01$.

## B. Additional Robustness Checks

Table B.1: Heterogeneous Treatment Effects Using 1.5x the Poverty Line.[a]

| Panel A (ITT): | Pre-Transfer Income <= 1.5*Poverty Line | | Pre-Transfer Income > 1.5*Poverty Line | |
|---|---|---|---|---|
| | Transfer From Children (Yes=1) | Transfer From Children (Yuan) | Transfer From Children (Yes=1) | Transfer From Children (Yuan) |
| | (1) | (2) | (3) | (4) |
| **OLS** | | | | |
| Offered NRPS* Above60 [b] | -0.029 | -174.960 | -0.068** | -618.438 |
| | (0.030) | (665.317) | (0.029) | (539.373) |
| Baseline Mean | 0.464 | 2041.368 | 0.319 | 1477.774 |
| Controls | Yes | Yes | Yes | Yes |
| R-squared | 0.259 | 0.085 | 0.248 | 0.057 |
| Observations | 3,635 | 3,627 | 4,449 | 4,441 |
| **Tobit [c]** | | | | |
| Marginal effect, E(T) | -0.063 | -672.061 | -0.143*** | -1959.077** |
| | (0.046) | (971.646) | (-0.048) | (-786.649) |
| Marginal effect, E[T \| T>0] | -0.04 | -250.749 | -0.073*** | -662.442** |
| | (0.029) | (355.5855) | (0.024) | (263.164) |
| Baseline Mean | 0.464 | 2041.368 | 0.319 | 1477.774 |
| Controls | Yes | Yes | Yes | Yes |
| Pseudo R-squared | 0.188 | 0.016 | 0.170 | 0.017 |
| Observations | 3,652 | 3,644 | 4,450 | 4,442 |
| **Panel B (TOT):** | | | | |
| **2SLS** | | | | |
| NRPS Receipt* Above60 [d] | -0.042 | -254.591 | -0.094** | -859.657 |
| | (0.043) | (968.908) | (0.041) | (749.976) |
| Baseline Mean | 0.464 | 2041.368 | 0.319 | 1477.774 |
| Controls | Yes | Yes | Yes | Yes |
| F-Stat (First Stage) | . | . | 419.86 | 420.41 |
| R-squared | 0.260 | 0.085 | 0.246 | 0.057 |
| Observations | 3,635 | 3,627 | 4,449 | 4,441 |
| **Tobit-IV [e]** | | | | |
| Marginal effect, E(T) | -0.093 | -979.501 | -0.199*** | -2719.265** |
| | (0.066) | (1396.671) | (0.066) | (1097.652) |
| Marginal effect, E[T \| T>0] | -0.058 | -363.368 | -0.103*** | -928.655** |
| | (0.041) | (505.541) | (0.035) | (374.146) |
| Baseline Mean | 0.319 | 1477.774 | 0.319 | 1477.774 |
| Controls | Yes | Yes | Yes | Yes |
| F-Stat (First Stage) | . | . | 546.98 | 546.98 |
| Pseudo R-squared | 0.187 | 0.015 | 0.170 | 0.017 |
| Observations | 3,635 | 3,627 | 4,449 | 4,441 |

Notes: a) Based on the World Bank's Income Classification for Millennium Development Goals, China classifies the poverty standard as per capita income of 2,300 in 2010 yuan (World Bank, 2017). (b) Policy instrument interacted with an indicator for being over 60 years old. (c) Tobit panel: Top row is the partial effect on latent dependent variable. Bottom row is the marginal effect on positive transfers (the effect of interest). (d) Endogenous NRPS Receipt is instrumented by a policy variable (Offer NRPS). (e) Tobit-IV panel: Top row is the partial effect on latent dependent variable. Bottom row is the marginal effect on positive transfers (the effect of interest). Individual level controls: Age, Age Squared, Marital Status (=1 if Married), Gender (=1 if Female), Education Levels (Base Group is illiterate with no formal education), Household Size (# of Household Residents). All regressions are estimated with Community, Year and (Community X Year) fixed effects. Clustered standard errors at the community level reported in parenthesis. *p< 0.10, **p< 0.05, ***p< 0.01.

Table B.2: Heterogeneous Treatment Effects Using 2x the Poverty Line.[a]

|  | Pre-Transfer Income <= 2*Poverty Line | | Pre-Transfer Income > 2*Poverty Line | |
|---|---|---|---|---|
| Panel A (ITT): | Transfer From Children (Yes=1) | Transfer From Children (Yuan) | Transfer From Children (Yes=1) | Transfer From Children (Yuan) |
|  | (1) | (2) | (3) | (4) |
| **OLS** | | | | |
| Offered NRPS * Above60 [b] | -0.042 | -441.830 | -0.067** | -405.015 |
|  | (0.025) | (565.126) | (0.031) | (622.075) |
| Baseline Mean | 0.450 | 1997.352 | 0.313 | 1440.244 |
| Controls | Yes | Yes | Yes | Yes |
| R-squared | 0.255 | 0.061 | 0.266 | 0.058 |
| Observations | 4,240 | 4,227 | 3,766 | 3,763 |
| **Tobit** [c] | | | | |
| Marginal effect, E(T) | -0.077* | -1034.035 | -0.151*** | -1893.169** |
|  | (0.040) | (844.277) | (0.051) | (880.318) |
| Marginal effect, E[T \| T>0] | -0.048* | -386.796 | -0.076*** | -630.324** |
|  | (0.025) | (305.828) | (0.025) | (289.924) |
| Baseline Mean | 0.450 | 1997.352 | 0.313 | 1440.244 |
| Controls | Yes | Yes | Yes | Yes |
| Pseudo R-squared | 0.181 | 0.014 | 0.187 | 0.019 |
| Observations | 4,258 | 4,245 | 3,771 | 3,768 |
| Panel B (TOT): | | | | |
| **2SLS** | | | | |
| NRPS Receipt* Above60 [d] | -0.061 | -664.727 | -0.093** | -563.317 |
|  | (0.043) | (826.728) | (0.043) | (865.708) |
| Baseline Mean | 0.450 | 1997.352 | 0.313 | 1440.244 |
| Controls | Yes | Yes | Yes | Yes |
| F-Stat (First Stage) | . | . | 362.43 | 362.03 |
| R-squared | 0.256 | 0.060 | 0.264 | 0.058 |
| Observations | 4,240 | 4,227 | 3,766 | 3,763 |
| **Tobit-IV** [e] | | | | |
| Marginal effect, E(T) | -0.113** | -1507.860 | -0.211*** | -2636.567** |
|  | (0.057) | (1223.803) | (0.071) | (1230.315) |
| Marginal effect, E[T \| T>0] | -0.070** | -559.160 | -0.107*** | -886.679** |
|  | (0.036) | (436.688) | (0.036) | (413.069) |
| Baseline Mean | 0.313 | 1440.244 | 0.313 | 1440.244 |
| Controls | Yes | Yes | Yes | Yes |
| F-Stat (First Stage) | . | . | 489.59 | 489.59 |
| Pseudo R-squared | 0.180 | 0.014 | 0.186 | 0.019 |
| Observations | 4,240 | 4,227 | 3,766 | 3,763 |

Notes: a) Based on the World Bank's Income Classification for Millennium Development Goals, China classifies the poverty standard as per capita income of 2,300 in 2010 yuan (World Bank, 2017). (b) Policy instrument interacted with an indicator for being over 60 years old. (c) Tobit panel: Top row is the partial effect on latent dependent variable. Bottom row is the marginal effect on positive transfers (the effect of interest). (d) Endogenous NRPS Receipt is instrumented by a policy variable (Offer NRPS). (e) Tobit-IV panel: Top row is the partial effect on latent dependent variable. Bottom row is the marginal effect on positive transfers (the effect of interest). Individual level controls: Age, Age Squared, Marital Status (=1 if Married), Gender (=1 if Female), Education Levels (Base Group is illiterate with no formal education), Household Size (# of Household Residents). All regressions are estimated with Community, Year and (Community X Year) fixed effects. Clustered standard errors at the community level reported in parenthesis. *p< 0.10, **p< 0.05, ***p< 0.01.